\documentclass[conference]{IEEEtran}
\IEEEoverridecommandlockouts
\usepackage{cite}
\usepackage{amsmath,amssymb,amsfonts}
\usepackage{algorithmic}
\usepackage{graphicx}
\usepackage{textcomp}
\usepackage{url}
\usepackage{booktabs} 
\usepackage[draft]{minted} 
\usepackage{subcaption}
\usepackage{textcomp}
\graphicspath{{figs/}}
\def\BibTeX{{\rm B\kern-.05em{\sc i\kern-.025em b}\kern-.08em
    T\kern-.1667em\lower.7ex\hbox{E}\kern-.125emX}}

\makeatletter
\def\PYG@reset{\let\PYG@it=\relax \let\PYG@bf=\relax%
	\let\PYG@ul=\relax \let\PYG@tc=\relax%
	\let\PYG@bc=\relax \let\PYG@ff=\relax}
\def\PYG@tok#1{\csname PYG@tok@#1\endcsname}
\def\PYG@toks#1+{\ifx\relax#1\empty\else%
	\PYG@tok{#1}\expandafter\PYG@toks\fi}
\def\PYG@do#1{\PYG@bc{\PYG@tc{\PYG@ul{%
				\PYG@it{\PYG@bf{\PYG@ff{#1}}}}}}}
\def\PYG#1#2{\PYG@reset\PYG@toks#1+\relax+\PYG@do{#2}}

\expandafter\def\csname PYG@tok@vc\endcsname{\def\PYG@tc##1{\textcolor[rgb]{0.10,0.09,0.49}{##1}}}
\expandafter\def\csname PYG@tok@gp\endcsname{\let\PYG@bf=\textbf\def\PYG@tc##1{\textcolor[rgb]{0.00,0.00,0.50}{##1}}}
\expandafter\def\csname PYG@tok@nd\endcsname{\def\PYG@tc##1{\textcolor[rgb]{0.67,0.13,1.00}{##1}}}
\expandafter\def\csname PYG@tok@cp\endcsname{\def\PYG@tc##1{\textcolor[rgb]{0.74,0.48,0.00}{##1}}}
\expandafter\def\csname PYG@tok@mb\endcsname{\def\PYG@tc##1{\textcolor[rgb]{0.40,0.40,0.40}{##1}}}
\expandafter\def\csname PYG@tok@kt\endcsname{\def\PYG@tc##1{\textcolor[rgb]{0.69,0.00,0.25}{##1}}}
\expandafter\def\csname PYG@tok@nl\endcsname{\def\PYG@tc##1{\textcolor[rgb]{0.63,0.63,0.00}{##1}}}
\expandafter\def\csname PYG@tok@vg\endcsname{\def\PYG@tc##1{\textcolor[rgb]{0.10,0.09,0.49}{##1}}}
\expandafter\def\csname PYG@tok@vi\endcsname{\def\PYG@tc##1{\textcolor[rgb]{0.10,0.09,0.49}{##1}}}
\expandafter\def\csname PYG@tok@cs\endcsname{\let\PYG@it=\textit\def\PYG@tc##1{\textcolor[rgb]{0.25,0.50,0.50}{##1}}}
\expandafter\def\csname PYG@tok@mo\endcsname{\def\PYG@tc##1{\textcolor[rgb]{0.40,0.40,0.40}{##1}}}
\expandafter\def\csname PYG@tok@ne\endcsname{\let\PYG@bf=\textbf\def\PYG@tc##1{\textcolor[rgb]{0.82,0.25,0.23}{##1}}}
\expandafter\def\csname PYG@tok@ss\endcsname{\def\PYG@tc##1{\textcolor[rgb]{0.10,0.09,0.49}{##1}}}
\expandafter\def\csname PYG@tok@sc\endcsname{\def\PYG@tc##1{\textcolor[rgb]{0.73,0.13,0.13}{##1}}}
\expandafter\def\csname PYG@tok@mf\endcsname{\def\PYG@tc##1{\textcolor[rgb]{0.40,0.40,0.40}{##1}}}
\expandafter\def\csname PYG@tok@mi\endcsname{\def\PYG@tc##1{\textcolor[rgb]{0.40,0.40,0.40}{##1}}}
\expandafter\def\csname PYG@tok@se\endcsname{\let\PYG@bf=\textbf\def\PYG@tc##1{\textcolor[rgb]{0.73,0.40,0.13}{##1}}}
\expandafter\def\csname PYG@tok@c\endcsname{\let\PYG@it=\textit\def\PYG@tc##1{\textcolor[rgb]{0.25,0.50,0.50}{##1}}}
\expandafter\def\csname PYG@tok@o\endcsname{\def\PYG@tc##1{\textcolor[rgb]{0.40,0.40,0.40}{##1}}}
\expandafter\def\csname PYG@tok@kp\endcsname{\def\PYG@tc##1{\textcolor[rgb]{0.00,0.50,0.00}{##1}}}
\expandafter\def\csname PYG@tok@w\endcsname{\def\PYG@tc##1{\textcolor[rgb]{0.73,0.73,0.73}{##1}}}
\expandafter\def\csname PYG@tok@gt\endcsname{\def\PYG@tc##1{\textcolor[rgb]{0.00,0.27,0.87}{##1}}}
\expandafter\def\csname PYG@tok@c1\endcsname{\let\PYG@it=\textit\def\PYG@tc##1{\textcolor[rgb]{0.25,0.50,0.50}{##1}}}
\expandafter\def\csname PYG@tok@gh\endcsname{\let\PYG@bf=\textbf\def\PYG@tc##1{\textcolor[rgb]{0.00,0.00,0.50}{##1}}}
\expandafter\def\csname PYG@tok@gd\endcsname{\def\PYG@tc##1{\textcolor[rgb]{0.63,0.00,0.00}{##1}}}
\expandafter\def\csname PYG@tok@ow\endcsname{\let\PYG@bf=\textbf\def\PYG@tc##1{\textcolor[rgb]{0.67,0.13,1.00}{##1}}}
\expandafter\def\csname PYG@tok@sd\endcsname{\let\PYG@it=\textit\def\PYG@tc##1{\textcolor[rgb]{0.73,0.13,0.13}{##1}}}
\expandafter\def\csname PYG@tok@k\endcsname{\let\PYG@bf=\textbf\def\PYG@tc##1{\textcolor[rgb]{0.00,0.50,0.00}{##1}}}
\expandafter\def\csname PYG@tok@gi\endcsname{\def\PYG@tc##1{\textcolor[rgb]{0.00,0.63,0.00}{##1}}}
\expandafter\def\csname PYG@tok@kc\endcsname{\let\PYG@bf=\textbf\def\PYG@tc##1{\textcolor[rgb]{0.00,0.50,0.00}{##1}}}
\expandafter\def\csname PYG@tok@gu\endcsname{\let\PYG@bf=\textbf\def\PYG@tc##1{\textcolor[rgb]{0.50,0.00,0.50}{##1}}}
\expandafter\def\csname PYG@tok@sb\endcsname{\def\PYG@tc##1{\textcolor[rgb]{0.73,0.13,0.13}{##1}}}
\expandafter\def\csname PYG@tok@nc\endcsname{\let\PYG@bf=\textbf\def\PYG@tc##1{\textcolor[rgb]{0.00,0.00,1.00}{##1}}}
\expandafter\def\csname PYG@tok@nb\endcsname{\def\PYG@tc##1{\textcolor[rgb]{0.00,0.50,0.00}{##1}}}
\expandafter\def\csname PYG@tok@bp\endcsname{\def\PYG@tc##1{\textcolor[rgb]{0.00,0.50,0.00}{##1}}}
\expandafter\def\csname PYG@tok@m\endcsname{\def\PYG@tc##1{\textcolor[rgb]{0.40,0.40,0.40}{##1}}}
\expandafter\def\csname PYG@tok@sh\endcsname{\def\PYG@tc##1{\textcolor[rgb]{0.73,0.13,0.13}{##1}}}
\expandafter\def\csname PYG@tok@no\endcsname{\def\PYG@tc##1{\textcolor[rgb]{0.53,0.00,0.00}{##1}}}
\expandafter\def\csname PYG@tok@gr\endcsname{\def\PYG@tc##1{\textcolor[rgb]{1.00,0.00,0.00}{##1}}}
\expandafter\def\csname PYG@tok@mh\endcsname{\def\PYG@tc##1{\textcolor[rgb]{0.40,0.40,0.40}{##1}}}
\expandafter\def\csname PYG@tok@gs\endcsname{\let\PYG@bf=\textbf}
\expandafter\def\csname PYG@tok@sr\endcsname{\def\PYG@tc##1{\textcolor[rgb]{0.73,0.40,0.53}{##1}}}
\expandafter\def\csname PYG@tok@kn\endcsname{\let\PYG@bf=\textbf\def\PYG@tc##1{\textcolor[rgb]{0.00,0.50,0.00}{##1}}}
\expandafter\def\csname PYG@tok@ni\endcsname{\let\PYG@bf=\textbf\def\PYG@tc##1{\textcolor[rgb]{0.60,0.60,0.60}{##1}}}
\expandafter\def\csname PYG@tok@kd\endcsname{\let\PYG@bf=\textbf\def\PYG@tc##1{\textcolor[rgb]{0.00,0.50,0.00}{##1}}}
\expandafter\def\csname PYG@tok@si\endcsname{\let\PYG@bf=\textbf\def\PYG@tc##1{\textcolor[rgb]{0.73,0.40,0.53}{##1}}}
\expandafter\def\csname PYG@tok@ch\endcsname{\let\PYG@it=\textit\def\PYG@tc##1{\textcolor[rgb]{0.25,0.50,0.50}{##1}}}
\expandafter\def\csname PYG@tok@cpf\endcsname{\let\PYG@it=\textit\def\PYG@tc##1{\textcolor[rgb]{0.25,0.50,0.50}{##1}}}
\expandafter\def\csname PYG@tok@nf\endcsname{\def\PYG@tc##1{\textcolor[rgb]{0.00,0.00,1.00}{##1}}}
\expandafter\def\csname PYG@tok@na\endcsname{\def\PYG@tc##1{\textcolor[rgb]{0.49,0.56,0.16}{##1}}}
\expandafter\def\csname PYG@tok@s1\endcsname{\def\PYG@tc##1{\textcolor[rgb]{0.73,0.13,0.13}{##1}}}
\expandafter\def\csname PYG@tok@nt\endcsname{\let\PYG@bf=\textbf\def\PYG@tc##1{\textcolor[rgb]{0.00,0.50,0.00}{##1}}}
\expandafter\def\csname PYG@tok@kr\endcsname{\let\PYG@bf=\textbf\def\PYG@tc##1{\textcolor[rgb]{0.00,0.50,0.00}{##1}}}
\expandafter\def\csname PYG@tok@ge\endcsname{\let\PYG@it=\textit}
\expandafter\def\csname PYG@tok@go\endcsname{\def\PYG@tc##1{\textcolor[rgb]{0.53,0.53,0.53}{##1}}}
\expandafter\def\csname PYG@tok@s2\endcsname{\def\PYG@tc##1{\textcolor[rgb]{0.73,0.13,0.13}{##1}}}
\expandafter\def\csname PYG@tok@nn\endcsname{\let\PYG@bf=\textbf\def\PYG@tc##1{\textcolor[rgb]{0.00,0.00,1.00}{##1}}}
\expandafter\def\csname PYG@tok@s\endcsname{\def\PYG@tc##1{\textcolor[rgb]{0.73,0.13,0.13}{##1}}}
\expandafter\def\csname PYG@tok@cm\endcsname{\let\PYG@it=\textit\def\PYG@tc##1{\textcolor[rgb]{0.25,0.50,0.50}{##1}}}
\expandafter\def\csname PYG@tok@sx\endcsname{\def\PYG@tc##1{\textcolor[rgb]{0.00,0.50,0.00}{##1}}}
\expandafter\def\csname PYG@tok@err\endcsname{\def\PYG@bc##1{\setlength{\fboxsep}{0pt}\fcolorbox[rgb]{1.00,0.00,0.00}{1,1,1}{\strut ##1}}}
\expandafter\def\csname PYG@tok@il\endcsname{\def\PYG@tc##1{\textcolor[rgb]{0.40,0.40,0.40}{##1}}}
\expandafter\def\csname PYG@tok@nv\endcsname{\def\PYG@tc##1{\textcolor[rgb]{0.10,0.09,0.49}{##1}}}

\def\PYGZus{\char`\_}

\def\PYGZgt{\char`\>}
\def\PYGZsh{\char`\#}

\def\PYGZdl{\char`\$}

\def\PYGZsq{\char`\'}
\def\PYGZdq{\char`\"}

\makeatother

\makeatletter
\def\PYGdefault@reset{\let\PYGdefault@it=\relax \let\PYGdefault@bf=\relax%
	\let\PYGdefault@ul=\relax \let\PYGdefault@tc=\relax%
	\let\PYGdefault@bc=\relax \let\PYGdefault@ff=\relax}
\def\PYGdefault@tok#1{\csname PYGdefault@tok@#1\endcsname}
\def\PYGdefault@toks#1+{\ifx\relax#1\empty\else%
	\PYGdefault@tok{#1}\expandafter\PYGdefault@toks\fi}
\def\PYGdefault@do#1{\PYGdefault@bc{\PYGdefault@tc{\PYGdefault@ul{%
				\PYGdefault@it{\PYGdefault@bf{\PYGdefault@ff{#1}}}}}}}
\def\PYGdefault#1#2{\PYGdefault@reset\PYGdefault@toks#1+\relax+\PYGdefault@do{#2}}

\expandafter\def\csname PYGdefault@tok@vg\endcsname{\def\PYGdefault@tc##1{\textcolor[rgb]{0.10,0.09,0.49}{##1}}}
\expandafter\def\csname PYGdefault@tok@vc\endcsname{\def\PYGdefault@tc##1{\textcolor[rgb]{0.10,0.09,0.49}{##1}}}
\expandafter\def\csname PYGdefault@tok@s\endcsname{\def\PYGdefault@tc##1{\textcolor[rgb]{0.73,0.13,0.13}{##1}}}
\expandafter\def\csname PYGdefault@tok@nf\endcsname{\def\PYGdefault@tc##1{\textcolor[rgb]{0.00,0.00,1.00}{##1}}}
\expandafter\def\csname PYGdefault@tok@kc\endcsname{\let\PYGdefault@bf=\textbf\def\PYGdefault@tc##1{\textcolor[rgb]{0.00,0.50,0.00}{##1}}}
\expandafter\def\csname PYGdefault@tok@bp\endcsname{\def\PYGdefault@tc##1{\textcolor[rgb]{0.00,0.50,0.00}{##1}}}
\expandafter\def\csname PYGdefault@tok@il\endcsname{\def\PYGdefault@tc##1{\textcolor[rgb]{0.40,0.40,0.40}{##1}}}
\expandafter\def\csname PYGdefault@tok@sb\endcsname{\def\PYGdefault@tc##1{\textcolor[rgb]{0.73,0.13,0.13}{##1}}}
\expandafter\def\csname PYGdefault@tok@nv\endcsname{\def\PYGdefault@tc##1{\textcolor[rgb]{0.10,0.09,0.49}{##1}}}
\expandafter\def\csname PYGdefault@tok@m\endcsname{\def\PYGdefault@tc##1{\textcolor[rgb]{0.40,0.40,0.40}{##1}}}
\expandafter\def\csname PYGdefault@tok@ge\endcsname{\let\PYGdefault@it=\textit}
\expandafter\def\csname PYGdefault@tok@nb\endcsname{\def\PYGdefault@tc##1{\textcolor[rgb]{0.00,0.50,0.00}{##1}}}
\expandafter\def\csname PYGdefault@tok@mi\endcsname{\def\PYGdefault@tc##1{\textcolor[rgb]{0.40,0.40,0.40}{##1}}}
\expandafter\def\csname PYGdefault@tok@gu\endcsname{\let\PYGdefault@bf=\textbf\def\PYGdefault@tc##1{\textcolor[rgb]{0.50,0.00,0.50}{##1}}}
\expandafter\def\csname PYGdefault@tok@c1\endcsname{\let\PYGdefault@it=\textit\def\PYGdefault@tc##1{\textcolor[rgb]{0.25,0.50,0.50}{##1}}}
\expandafter\def\csname PYGdefault@tok@sd\endcsname{\let\PYGdefault@it=\textit\def\PYGdefault@tc##1{\textcolor[rgb]{0.73,0.13,0.13}{##1}}}
\expandafter\def\csname PYGdefault@tok@s1\endcsname{\def\PYGdefault@tc##1{\textcolor[rgb]{0.73,0.13,0.13}{##1}}}
\expandafter\def\csname PYGdefault@tok@nn\endcsname{\let\PYGdefault@bf=\textbf\def\PYGdefault@tc##1{\textcolor[rgb]{0.00,0.00,1.00}{##1}}}
\expandafter\def\csname PYGdefault@tok@ss\endcsname{\def\PYGdefault@tc##1{\textcolor[rgb]{0.10,0.09,0.49}{##1}}}
\expandafter\def\csname PYGdefault@tok@sc\endcsname{\def\PYGdefault@tc##1{\textcolor[rgb]{0.73,0.13,0.13}{##1}}}
\expandafter\def\csname PYGdefault@tok@nl\endcsname{\def\PYGdefault@tc##1{\textcolor[rgb]{0.63,0.63,0.00}{##1}}}
\expandafter\def\csname PYGdefault@tok@nc\endcsname{\let\PYGdefault@bf=\textbf\def\PYGdefault@tc##1{\textcolor[rgb]{0.00,0.00,1.00}{##1}}}
\expandafter\def\csname PYGdefault@tok@ni\endcsname{\let\PYGdefault@bf=\textbf\def\PYGdefault@tc##1{\textcolor[rgb]{0.60,0.60,0.60}{##1}}}
\expandafter\def\csname PYGdefault@tok@gi\endcsname{\def\PYGdefault@tc##1{\textcolor[rgb]{0.00,0.63,0.00}{##1}}}
\expandafter\def\csname PYGdefault@tok@w\endcsname{\def\PYGdefault@tc##1{\textcolor[rgb]{0.73,0.73,0.73}{##1}}}
\expandafter\def\csname PYGdefault@tok@cs\endcsname{\let\PYGdefault@it=\textit\def\PYGdefault@tc##1{\textcolor[rgb]{0.25,0.50,0.50}{##1}}}
\expandafter\def\csname PYGdefault@tok@ne\endcsname{\let\PYGdefault@bf=\textbf\def\PYGdefault@tc##1{\textcolor[rgb]{0.82,0.25,0.23}{##1}}}
\expandafter\def\csname PYGdefault@tok@go\endcsname{\def\PYGdefault@tc##1{\textcolor[rgb]{0.53,0.53,0.53}{##1}}}
\expandafter\def\csname PYGdefault@tok@mb\endcsname{\def\PYGdefault@tc##1{\textcolor[rgb]{0.40,0.40,0.40}{##1}}}
\expandafter\def\csname PYGdefault@tok@kd\endcsname{\let\PYGdefault@bf=\textbf\def\PYGdefault@tc##1{\textcolor[rgb]{0.00,0.50,0.00}{##1}}}
\expandafter\def\csname PYGdefault@tok@sh\endcsname{\def\PYGdefault@tc##1{\textcolor[rgb]{0.73,0.13,0.13}{##1}}}
\expandafter\def\csname PYGdefault@tok@sx\endcsname{\def\PYGdefault@tc##1{\textcolor[rgb]{0.00,0.50,0.00}{##1}}}
\expandafter\def\csname PYGdefault@tok@na\endcsname{\def\PYGdefault@tc##1{\textcolor[rgb]{0.49,0.56,0.16}{##1}}}
\expandafter\def\csname PYGdefault@tok@mf\endcsname{\def\PYGdefault@tc##1{\textcolor[rgb]{0.40,0.40,0.40}{##1}}}
\expandafter\def\csname PYGdefault@tok@kr\endcsname{\let\PYGdefault@bf=\textbf\def\PYGdefault@tc##1{\textcolor[rgb]{0.00,0.50,0.00}{##1}}}
\expandafter\def\csname PYGdefault@tok@cm\endcsname{\let\PYGdefault@it=\textit\def\PYGdefault@tc##1{\textcolor[rgb]{0.25,0.50,0.50}{##1}}}
\expandafter\def\csname PYGdefault@tok@no\endcsname{\def\PYGdefault@tc##1{\textcolor[rgb]{0.53,0.00,0.00}{##1}}}
\expandafter\def\csname PYGdefault@tok@kt\endcsname{\def\PYGdefault@tc##1{\textcolor[rgb]{0.69,0.00,0.25}{##1}}}
\expandafter\def\csname PYGdefault@tok@c\endcsname{\let\PYGdefault@it=\textit\def\PYGdefault@tc##1{\textcolor[rgb]{0.25,0.50,0.50}{##1}}}
\expandafter\def\csname PYGdefault@tok@gd\endcsname{\def\PYGdefault@tc##1{\textcolor[rgb]{0.63,0.00,0.00}{##1}}}
\expandafter\def\csname PYGdefault@tok@cpf\endcsname{\let\PYGdefault@it=\textit\def\PYGdefault@tc##1{\textcolor[rgb]{0.25,0.50,0.50}{##1}}}
\expandafter\def\csname PYGdefault@tok@se\endcsname{\let\PYGdefault@bf=\textbf\def\PYGdefault@tc##1{\textcolor[rgb]{0.73,0.40,0.13}{##1}}}
\expandafter\def\csname PYGdefault@tok@gp\endcsname{\let\PYGdefault@bf=\textbf\def\PYGdefault@tc##1{\textcolor[rgb]{0.00,0.00,0.50}{##1}}}
\expandafter\def\csname PYGdefault@tok@si\endcsname{\let\PYGdefault@bf=\textbf\def\PYGdefault@tc##1{\textcolor[rgb]{0.73,0.40,0.53}{##1}}}
\expandafter\def\csname PYGdefault@tok@sr\endcsname{\def\PYGdefault@tc##1{\textcolor[rgb]{0.73,0.40,0.53}{##1}}}
\expandafter\def\csname PYGdefault@tok@gr\endcsname{\def\PYGdefault@tc##1{\textcolor[rgb]{1.00,0.00,0.00}{##1}}}
\expandafter\def\csname PYGdefault@tok@gh\endcsname{\let\PYGdefault@bf=\textbf\def\PYGdefault@tc##1{\textcolor[rgb]{0.00,0.00,0.50}{##1}}}
\expandafter\def\csname PYGdefault@tok@o\endcsname{\def\PYGdefault@tc##1{\textcolor[rgb]{0.40,0.40,0.40}{##1}}}
\expandafter\def\csname PYGdefault@tok@kp\endcsname{\def\PYGdefault@tc##1{\textcolor[rgb]{0.00,0.50,0.00}{##1}}}
\expandafter\def\csname PYGdefault@tok@err\endcsname{\def\PYGdefault@bc##1{\setlength{\fboxsep}{0pt}\fcolorbox[rgb]{1.00,0.00,0.00}{1,1,1}{\strut ##1}}}
\expandafter\def\csname PYGdefault@tok@mo\endcsname{\def\PYGdefault@tc##1{\textcolor[rgb]{0.40,0.40,0.40}{##1}}}
\expandafter\def\csname PYGdefault@tok@k\endcsname{\let\PYGdefault@bf=\textbf\def\PYGdefault@tc##1{\textcolor[rgb]{0.00,0.50,0.00}{##1}}}
\expandafter\def\csname PYGdefault@tok@gs\endcsname{\let\PYGdefault@bf=\textbf}
\expandafter\def\csname PYGdefault@tok@ow\endcsname{\let\PYGdefault@bf=\textbf\def\PYGdefault@tc##1{\textcolor[rgb]{0.67,0.13,1.00}{##1}}}
\expandafter\def\csname PYGdefault@tok@mh\endcsname{\def\PYGdefault@tc##1{\textcolor[rgb]{0.40,0.40,0.40}{##1}}}
\expandafter\def\csname PYGdefault@tok@cp\endcsname{\def\PYGdefault@tc##1{\textcolor[rgb]{0.74,0.48,0.00}{##1}}}
\expandafter\def\csname PYGdefault@tok@nd\endcsname{\def\PYGdefault@tc##1{\textcolor[rgb]{0.67,0.13,1.00}{##1}}}
\expandafter\def\csname PYGdefault@tok@nt\endcsname{\let\PYGdefault@bf=\textbf\def\PYGdefault@tc##1{\textcolor[rgb]{0.00,0.50,0.00}{##1}}}
\expandafter\def\csname PYGdefault@tok@vi\endcsname{\def\PYGdefault@tc##1{\textcolor[rgb]{0.10,0.09,0.49}{##1}}}
\expandafter\def\csname PYGdefault@tok@s2\endcsname{\def\PYGdefault@tc##1{\textcolor[rgb]{0.73,0.13,0.13}{##1}}}
\expandafter\def\csname PYGdefault@tok@ch\endcsname{\let\PYGdefault@it=\textit\def\PYGdefault@tc##1{\textcolor[rgb]{0.25,0.50,0.50}{##1}}}
\expandafter\def\csname PYGdefault@tok@kn\endcsname{\let\PYGdefault@bf=\textbf\def\PYGdefault@tc##1{\textcolor[rgb]{0.00,0.50,0.00}{##1}}}
\expandafter\def\csname PYGdefault@tok@gt\endcsname{\def\PYGdefault@tc##1{\textcolor[rgb]{0.00,0.27,0.87}{##1}}}

\makeatother

\begin{document}

\title{Data Analytics Service Composition and Deployment on Edge Devices}

\author{\IEEEauthorblockN{Jianxin Zhao, Tudor Tiplea, Richard Mortier, Jon Crowcroft, Liang Wang}
\IEEEauthorblockA{Computer Laboratory, University of Cambridge, UK \\
firstname.lastname@cl.cam.ac.uk}
}

\maketitle

\begin{abstract}
	Data analytics on edge devices has gained rapid growth in research, industry, and different aspects of our daily life.
	This topic still faces many challenges such as limited computation resource on edge devices.
	In this paper, we further identify two main challenges: the composition and deployment of data analytics services on edge devices.
	We present the Zoo system to address these two challenge: on one hand, it provides simple and concise domain-specific language to enable easy and and type-safe composition of different data analytics services; on the other, it utilises multiple deployment backends, including Docker container, JavaScript, and MirageOS, to accommodate the heterogeneous edge deployment environment.
	We show the expressiveness of Zoo with a use case, and thoroughly compare the performance of different deployment backends in evaluation.
\end{abstract}

\section{Introduction}

Machine Learning (ML) techniques have begun to dominate data analytics applications and services.
Recommendation systems are the driving force of online service providers such as Amazon.
Finance analytics has quickly adopted ML to harness large volume of data in such areas as fraud detection and risk-management.
Deep Neural Network (DNN) is the technology behind voice-based personal assistance, self-driving cars~\cite{bojarski2016end}, image processing~\cite{googlevisionapi}, \emph{etc}.
Many popular data analytics are deployed on cloud computing infrastructures.
However, they require aggregating users' data at central server for processing. This architecture is prone to issues such as increased service response latency, communication cost, single point failure, and data privacy concerns.

Recently computation on edge and mobile devices has gained rapid growth, 
such as personal data analytics in home~\cite{mortier2016personal}, DNN application on a tiny stick~\cite{movidius17}, and semantic search and recommendation on web browser~\cite{wang2016kvasir}.
HUAWEI has identified speed and responsiveness of native AI processing on mobile devices as the key to a new era in smartphone innovation~\cite{huaweiai2012}.

Many challenges arise when moving ML analytics from cloud to edge devices.
One widely discussed challenge is the limited computation power and working memory of edge and mobile devices.
Personalising analytics models on different edge devices is also a very interesting topic~\cite{rodriguezWZMH17}.
However, one problem is not yet well defined and investigated: the deployment of data analytics services.
Most existing machine learning frameworks such as TensorFlow and Caffe focus mainly on the training of analytics models.
On the other, the end users, many of whom are not ML professionals, mainly use trained models to perform inference.
This gap between the current ML systems and users' requirements is growing.

Another challenge in conducting ML based data analytics on edge devices is model composition.
Training a model often requires large datasets and rich computing resources, which are often not available to normal users. That's one of the reasons that they are bounded with the models and services provided by large companies.
To this end we propose the idea \textit{Composable Service}.
Its basic idea is that 
many services can be constructed from basic ML ones such as image recognition, speech-to-text, and recommendation to meet new application requirements.
We believe that modularity and composition will be the key to increasing usage of ML-based data analytics.

This paper tries to address these two challenges. Specifically, the contribution of this paper includes:
\begin{itemize}
	\item We identify two challenges that are not yet well explored in the literature about data analytics on edge devices: service composition and deployment.
	\item We present the design of the Zoo system to address the previous two challenges. It provides concise Domain-specific Language (DSL) to enable composition of different data analytics services, and also deploys services to multiple backends.
	\item We present a use case to demonstrate the expressiveness of the DSL, and thoroughly evaluate different deployment backend for analytics services.
\end{itemize}

\section{Workflow}

\begin{figure}[!t]
	\centering
	\includegraphics[width=\columnwidth]{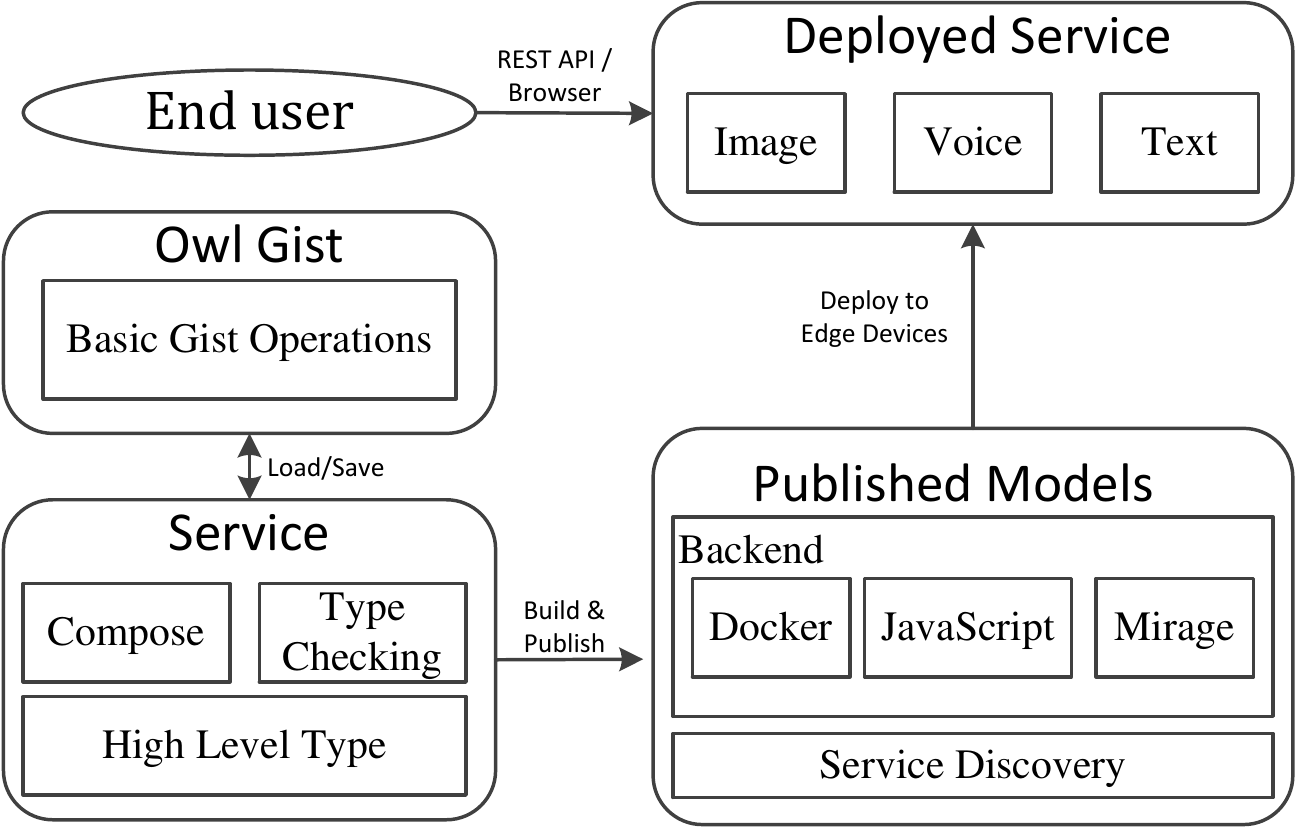}
	\caption{Zoo System Architecture}
	\label{fig:workflow}
	\vspace{-1em}
\end{figure}

Before presenting the system design, we would like to briefly introduce the workflow of Zoo as shown in Fig.~\ref{fig:workflow}. The workflow consists of two parts: \textit{development} on the left side and \textit{deployment} on the right.

\textit{Development} concerns the design of interaction workflow and the computational functions of different services.
One basic component is Github Gist.
A normal Gist script will be loaded as a module in OCaml.
To compose functionalities from different Gists only requires a developer to add one configuration file to each Gist.
This file is in JSON format. It consists of one or more name-value pairs. Each pair is a signature for a function the script developer wants to expose as a service.
These Gists then can be imported and composed to make new services.
When a user is satisfied with the composing result, she can save the new service as another Zoo Gist.

\textit{Deployment} takes a Gist and creates models in different backends. These models can be published and deployed to edge devices.
It is separated from the logic of development.
Basic services and composed ones are treated equally.
Besides, users can move services from being local to remote and vice versa, without changing the structure of the constructed service.
Deployment is not limited to edge devices, but can also be on cloud servers, or a hybrid of both cases, to minimise the data revealed to the cloud and the associated communication costs.
Thus by this design a data analytics service can easily be distributed to multiple devices.

\section{System Design}

The Zoo system is implemented on Owl~\cite{liang2017owl}, an open-source scientific computing library in OCaml language.
The reason we choose Owl to support the implementation of Zoo is some of its nice features.
Owl provides a full stack support for numerical methods, scientific computing, and advanced data analytics on OCaml.
Built on the core data structure of N-dimensional array (ndarray), Owl supports a comprehensive set of classic analytics such as math functions, statistics, linear algebra, as well as advanced analytics techniques, namely optimisation, algorithmic differentiation, and regression.
On top of them, Owl provides Neural Network and Natural Language Processing modules. Zoo relies on these modules to construct basic ML services.
OCaml provides static type checking, and Owl's ML modules have shown great expressiveness and code flexibility.

Initially, the Zoo system is designed to make it convenient for developers to share their OCaml code snippets. The design principle is to make the whole ecosystem open, flexible, and extensible.
One typical scenario for using the basic functions of Zoo can be described as follows. Developer A creates a script, uploads it to Gist, and then share it using a string of Gist id. When developer B gets this id, he can use the functions from A's scripts by simply using the ``\#zoo'' directive in his code. All the OCaml files in the Gist will be imported as modules for B to use.
Based on these basic functionalities, we'll explain how we extend the Zoo system to address the composition and deployment challenges.

\subsection{Service}
\label{subs:service}
Gist is a core abstraction in Zoo. It is the centre of code sharing.
However, to compose multiple analytics snippets, Gist alone is insufficient. For example, it cannot express the structure of how different pieces of code are composed together.
Therefore, we introduce another abstraction: \texttt{service}.

A service consists of three parts: \textit{Gists}, \textit{types}, and \textit{dependency graph}.
\textit{Gists} is the list of Gist ids this service requires.
\textit{Types} is the parameter types of this service. Any service has zero or more input parameters and one output. This design follows that of an OCaml function.
\textit{Dependency graph} is a graph structure that contains information about how the service is composed. Each node in it represents a function from a Gist, and contains the Gist's name, id, and number of parameters of this function.

Zoo provides three core operations about a service: create, compose, and publish.
The \textit{create\_service} creates a dictionary of services given a Gist id. This operation reads the service configuration file from that Gist, and creates a service for each function specified in the configuration file.
The \textit{compose\_service} provides a series of operations to combine multiple services into a new service. A compose operation does type checking by comparing the ``types'' field of two services. An error will be raised if incompatible services are composed. A composed service can be saved to a new Gist or be used for further composition.
The \textit{publish\_service} makes a service's code into such forms that can be readily used by end users. Zoo is designed to support multiple backends for these publication forms. Currently it targets Docker container, JavaScript, and MirageOS~\cite{madhavapeddy2013unikernels} as backends.

\subsection{Type Checking}

As mentioned in Section~\ref{subs:service}, one of the most important tasks of service composition is to make sure the type matches.
For example, suppose there is an image analytics service that takes a PNG format image, and if we connect to it another one that produces a JPEG image, the resulting service will only generate meaningless output for data type mismatch.
OCaml provides primary types such as integer, float, string, and bool.
The core data structure of Owl is ndarray (or tensor as it is called in some other data analytics frameworks).
However, all these types are insufficient for high level service type checking as mentioned. That motives us to derive richer high-level types.

To support it, we use generalised algebraic data types (GADTs) in OCaml.
There already exist several model collections on different platforms, e.g. Caffe~\cite{caffezoo2017} and MxNet~\cite{mxnetzoo2017}.
We observe that most current popular deep learning (DL) models can generally be categorised into three fundamental types: \texttt{image}, \texttt{text}, and \texttt{voice}. 
Based on them, we define sub-types for each: PNG and JPEG image, French and English text and voice, i.e. \texttt{png img}, \texttt{jpeg img}, \texttt{fr text}, \texttt{en text}, \texttt{fr voice}, and \texttt{en voice} types.
More can be further added easily in Zoo.
Therefore type checking in OCaml ensures type-safe and meaningful composition of high level services.

\subsection{Backend}

Recognising the heterogeneity of edge device deployment, one key principle of Zoo is to support multiple deployment methods.
Containerisation as a lightweight virtualisation technology has gained enormous traction. It is used in deployment systems such as Kubernetes.
Zoo supports deploying services as Docker containers. Each container provides RESTful API for end users to query.

Another backend is JavaScript.
Using JavaScript to do analytics aside from front end development begins to attract interests from academia~\cite{wang2016kvasir} and industry, such as Tensorflow.js and Facebook's Reason language~\cite{reasonml2017}.
By exporting OCaml and Owl functions to JavaScript code, users can do complex data analytics on web browser directly without relying on any other dependencies.

Aside from these two backends, we also initially explore using MirageOS as an option. Mirage is an example of Unikernel, which builds tiny virtual machines with a specialised minimal OS that host only one target application.
Deploying to Unikernel is proved to be of low memory footprint, and thus quite suitable for resource-limited edge devices.

\subsection{DSL}

Zoo provides a minimal DSL for service composition and deployment.

\paragraph{Composition}
To acquire services from a Gist of id \textit{gid}, we use $\$gid$ to create a dictionary, which maps from service name strings to services.
We implement the dictionary data structure using \texttt{Hashtbl} in OCaml.
The $\#$ operator is overloaded to represent the ``get item'' operation.
Therefore, \[\$\textrm{gid} \# \textrm{sname}\] can be used to get a service that is named ``sname''.
Now suppose we have $n$ services: $f_1$, $f_2$, \ldots, $f_n$. Their outputs are of type $t_{f1}$, $t_{f2}$, \ldots, $t_{fn}$. Each service $s$ accepts $m_s$ input parameters, which have type $t_s^1$, $t_s^2$, \ldots, $t_s^{m_s}$.
Also, there is a service $g$ that takes $n$ inputs, each of them has type  $t_g^1$, $t_g^2$, \ldots, $t_g^n$. Its output type is $t_o$.
Here Zoo provides the \texttt{\$>} operator to compose a list of services with another:
\[ [f_1, f_2, \ldots, f_n] \textrm{\$>} g \]
This operation returns a new service that has $\sum_{s=1}^{n} m_s$ inputs, and is of output type $t_o$. This operation does type checking to make sure that $ t_{fi} = t_g^i, \forall i \in {1, 2, \ldots, n}$.

\paragraph{Deployment} Taking a service $s$, be it a basic or composed one, it can be deployed using the following syntax:

\[ s \textrm{\$@ backend} \]

The \texttt{\$@} operator publish services to certain backend. It returns a string of URI of the resources to be deployed.

\subsection{Service Discovery}

The services requires a service discovery mechanism. For simplicity's sake, each newly published service is added to a public record hosted on a server.
The record is a list of items, and each item contains the Gist id that service based on, a one-line description of this service, string representation of the input types and output type of this service, e.g. ``image $\rightarrow$ int $\rightarrow$ string $\rightarrow$ tex'', and service URI. For the container deployment, the URI is a DockerHub link, and for JavaScript backend, the URI is a URL link to the JavaScript file itself.
The service discovery mechanism is implemented using off-the-shelf database.


\subsection{Version Control}

Developers would modify and upload their scripts several times. As such, each version of a script is assigned a unique id in Gist. Zoo supports specifying a version of a Gist.

The naming scheme of a Gist is \texttt{gid/[vid|latest]/pin}. A user can either choose a specific version id, or he can use the latest version, which means the newest version on local cache.
Obviously, ``latest'' will introduce cache inconsistency. The latest version on one machine might not be the same on the other. 
To get the up-to-date version from Gist server,
the download time of the latest version on a local machine will be saved as metadata. The newest version on server will be pulled to local cache after a certain period of time, if ``latest'' flag is set in the Gist name.
Ideally, every published service should contain a specific version id, and ``latest'' should only be used during development.

Zoo can analyse dependency information of a Gist and save it.
When the ``pin'' flag is set, Gist dependency graph of current script will be saved or loaded.

\section{Use Case}

To illustrate the workflow above, let's consider a synthetic scenario.
Alice is a French data analyst. She knows how to use ML and DL models in existing platforms, but is not an expert.
Her recent work is about testing the performance of different image classification neural networks. To do that, she need to first modify the image using the DNN-based Neural Style Transfer (NST) algorithm.
The NST algorithm takes two images and outputs to a new image, which is similar to the first image in content and the second in style.
This new image should be passed to an image classification DNN for inference.
Finally, the classification result should be translated to French.
She does not want to put academic-related information on Google's server, but she cannot find any single pre-trained model that performs this series of tasks.

Here comes the Zoo system to help. Alice find Gists that can do image recognition, NST, and translation separately.
Even better, she can perform image segmentation to greatly improve the performance of NST~\cite{LuanPSB17} using another Gist.
All she has to provide is some simple code to generate the style images she need to use.
She can then assemble these parts together easily using Zoo.

%

\begin{Verbatim}[commandchars=\\\{\},codes={\catcode`\$=3\catcode`\^=7\catcode`\_=8}]
\PYG{k}{open} \PYG{n+nc}{Zoo}
\PYG{c}{(* Image classification *)}
\PYG{k}{let} \PYG{n}{s\PYGZus{}img} \PYG{o}{=} \PYG{o}{\PYGZdl{}} \PYG{l+s+s2}{\PYGZdq{}aa36e\PYGZdq{}} \PYG{o}{\PYGZsh{}} \PYG{l+s+s2}{\PYGZdq{}infer\PYGZdq{}}\PYG{o}{;;}
\PYG{c}{(* Image segmentation *)}
\PYG{k}{let} \PYG{n}{s\PYGZus{}seg} \PYG{o}{=} \PYG{o}{\PYGZdl{}} \PYG{l+s+s2}{\PYGZdq{}d79e9\PYGZdq{}} \PYG{o}{\PYGZsh{}} \PYG{l+s+s2}{\PYGZdq{}seg\PYGZdq{}}\PYG{o}{;;}
\PYG{c}{(* Neural style transfer *)}
\PYG{k}{let} \PYG{n}{s\PYGZus{}nst} \PYG{o}{=} \PYG{o}{\PYGZdl{}} \PYG{l+s+s2}{\PYGZdq{}6f28d\PYGZdq{}} \PYG{o}{\PYGZsh{}} \PYG{l+s+s2}{\PYGZdq{}run\PYGZdq{}}\PYG{o}{;;}
\PYG{c}{(* Translation from English to French *)}
\PYG{k}{let} \PYG{n}{s\PYGZus{}trans} \PYG{o}{=} \PYG{o}{\PYGZdl{}} \PYG{l+s+s2}{\PYGZdq{}7f32a\PYGZdq{}} \PYG{o}{\PYGZsh{}} \PYG{l+s+s2}{\PYGZdq{}trans\PYGZdq{}}\PYG{o}{;;}
\PYG{c}{(* Alice\PYGZsq{}s image generation service *)}
\PYG{k}{let} \PYG{n}{s\PYGZus{}style} \PYG{o}{=} \PYG{o}{\PYGZdl{}} \PYG{n}{alice\PYGZus{}Gist\PYGZus{}id}
\PYG{o}{\PYGZsh{}} \PYG{l+s+s2}{\PYGZdq{}image\PYGZus{}gen\PYGZdq{}}\PYG{o}{;;}

\PYG{c}{(* Compose services *)}
\PYG{k}{let} \PYG{n}{s} \PYG{o}{=} \PYG{o}{[}\PYG{n}{s\PYGZus{}seg}\PYG{o}{;} \PYG{n}{s\PYGZus{}style}\PYG{o}{]} \PYG{o}{\PYGZdl{}\PYGZgt{}} \PYG{n}{s\PYGZus{}nst}
\PYG{o}{\PYGZdl{}\PYGZgt{}} \PYG{n}{n\PYGZus{}img} \PYG{o}{\PYGZdl{}\PYGZgt{}} \PYG{n}{n\PYGZus{}trans}\PYG{o}{;;}
\PYG{c}{(* Publish to a new Docker Image *)}
\PYG{k}{let} \PYG{n}{pub} \PYG{o}{=} \PYG{o}{(}\PYG{n+nn}{List}\PYG{p}{.}\PYG{n}{hd} \PYG{n}{s}\PYG{o}{)} \PYG{o}{\PYGZdl{}@}
\PYG{o}{(}\PYG{n+nc}{CONTAINER} \PYG{l+s+s2}{\PYGZdq{}alice/image\PYGZus{}service:latest\PYGZdq{}}\PYG{o}{);;}
\end{Verbatim}

Note that the Gist id used in the code is shorted from 32 digits to 5 due to column length limit.
Once Alice creates the news service and published it as a container, 
she can then run it locally and send request with image data to the deployed machine, and get image classification results back in French.

\section{Evaluation}
In the evaluation section we focus on comparing the performance of different backends we use.
Specifically, we observe three representative groups of operations:
(1) \texttt{map} and \texttt{fold} operations on ndarray;
(2) using gradient descent, a common numerical computing subroutine, to get $argmin$ of a certain function;
(3) conducting inference on complex DNNs, including SqueezeNet~\cite{iandola2016squeezenet} and a VGG-like convolution network.
The evaluations are conducted on a ThinkPad T460S laptop with Ubuntu 16.04 operating system. It has an Intel Core i5-6200U CPU and 12GB RAM.

The OCaml compiler can produce two kinds of executables: bytecode and native. Native executables are compiled specifically for an architecture and are generally faster, while bytecode executables have the advantage of being portable.
A Docker container can adopt both options.

For JavaScript though, since the Owl library contains functions that are implemented in C, it cannot be directly supported by \texttt{js-of-ocaml}, the tool we use to convert OCaml code into JavaScript.
Therefore in the Owl library, we have implemented a ``base'' library in pure OCaml that shares the core functions of the Owl library.
Note that for convenience we refer to the pure implementation of OCaml and the mix implementation of OCaml and C as \texttt{base-lib} and \texttt{owl-lib} separately, but they are in fact all included in the Owl library.
For Mirage compilation, we use both libraries.

\begin{figure*}[!t]
	\centering
	\includegraphics[width=\textwidth]{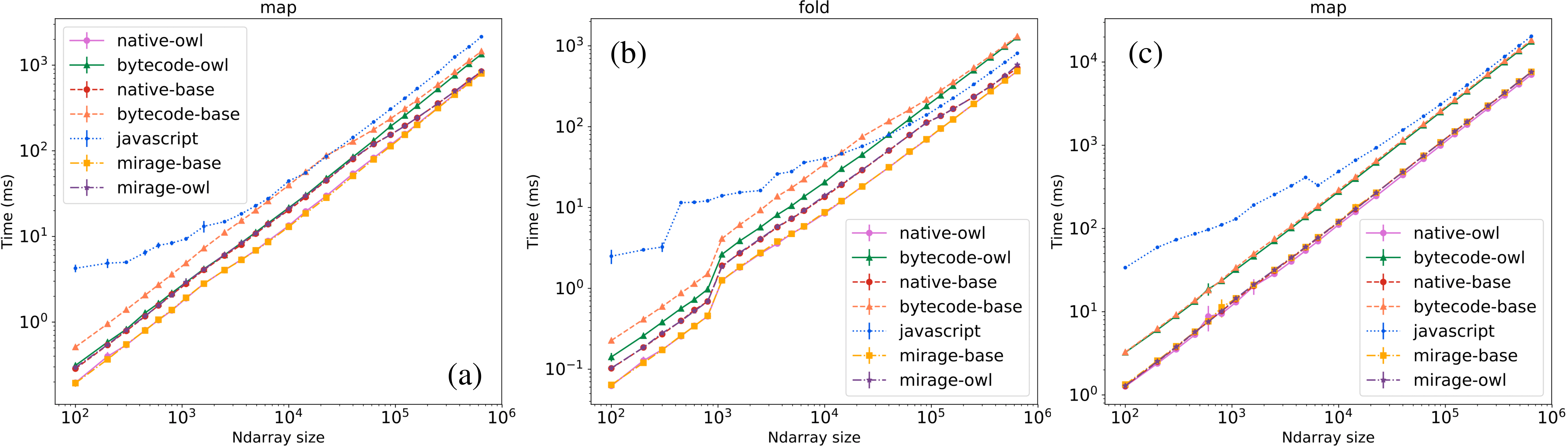}
	\caption{Performance of map and fold operations on ndarray on laptop (a-b) and RaspberryPi (c).}
	\label{fig:eval_mapfold}
	\vspace{-0.5em}
\end{figure*}

Fig.~\ref{fig:eval_mapfold}(a-b) show the performance of map and fold operations on ndarray. We use simple functions such as plus and multiplication on 1-d (size $< 1,000$ ) and 2-d arrays.
The log-log relationship between total size of ndarray and the time each operation takes keeps linear.
For both operations, \texttt{owl-lib} is faster than \texttt{base-lib}, and native executables outperform bytecode ones.
The performance of Mirage executives is close to that of native code.
Generally JavaScript runs the slowest, but note how the performance gap between JavaScript and the others converges when the ndarray size grows. For fold operation, JavaScript even runs faster than bytecode when size is sufficiently large.

\begin{figure}[!t]
	\centering
	\includegraphics[width=\columnwidth]{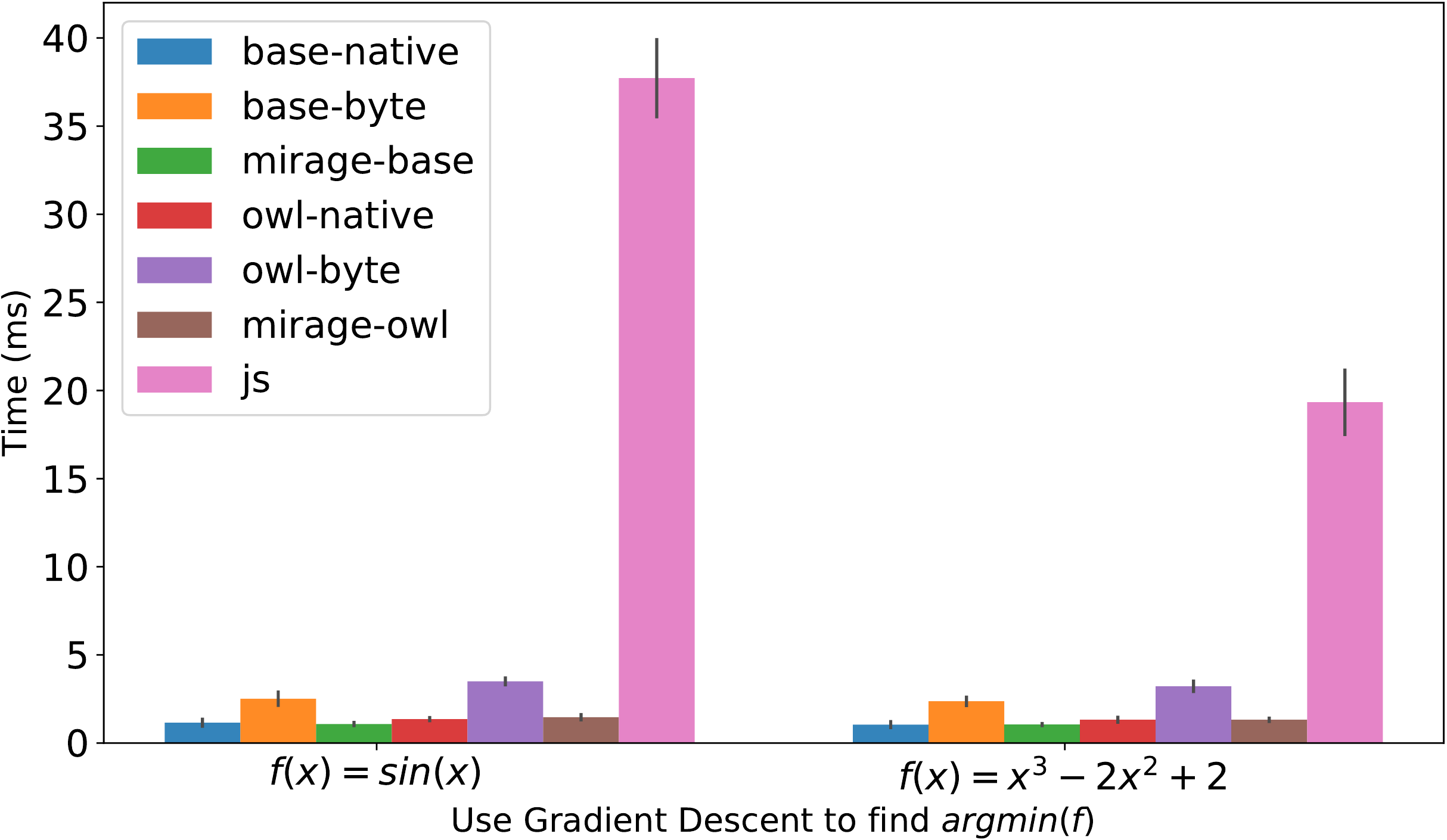}
	\caption{Performance of gradient descent on function $f$ to find $argmin(f)$ on laptop.}
	\label{fig:eval_gd}
\end{figure}

In Fig.~\ref{fig:eval_gd}, we want to investigate if the above observations still hold in more complex numerical computation.
We choose to use a Gradient Descent algorithm to find the value that locally minimise a function. We choose the initial value randomly between $[0, 10]$.
For both $sin(x)$ and $x^3 -2x^2 + 2$, we can see that JavaScript runs the slowest, but this time the \texttt{base-lib} slightly outperforms \texttt{owl-lib}.

\begin{table}
	\caption{Inference Speed of DNN (Laptop)}
	\label{tab:eval_dnn}
	\begin{tabular}{r|ll}
		\toprule
		Time (ms) &VGG&SqueezeNet\\
		\midrule
		owl-native & 7.96 ($\pm$ 0.93) & 196.26($\pm$ 1.12) \\
		owl-byte & 9.87 ($\pm$ 0.74) & 218.99($\pm$ 9.05) \\
		base-native & 792.56($\pm$ 19.95) & 14470.97 ($\pm$ 368.03)\\
		base-byte & 2783.33($\pm$ 76.08) & 50294.93 ($\pm$ 1315.28)\\
		mirage-owl & 8.09($\pm$ 0.08) & 190.26($\pm$ 0.89)\\
		mirage-base & 743.18 ($\pm$ 13.29) & 13478.53 ($\pm$ 13.29)\\
		JavaScript & 4325.50($\pm$ 447.22) & 65545.75 ($\pm$ 629.10)\\
		\bottomrule
	\end{tabular}
\end{table}

We further compare the performance of DNN, which requires large amount of computation.
We compare SqueezeNet and a VGG-like convolution network. They have different sizes of weight and networks structure complexities.
Table.~\ref{tab:eval_dnn} shows that, though the performance difference between \texttt{owl-lib} and \texttt{base-lib} is not obvious, the former is much better. So is the difference between native and bytecode for \texttt{base-lib}. JavaScript is still the slowest.
The core computation required for DNN inference is the convolution operation. Its implementation efficiency is the key to these differences.
Current we are working on improving its implementation in \texttt{base-lib}.


\begin{figure}[!t]
	\centering
	\includegraphics[width=\columnwidth]{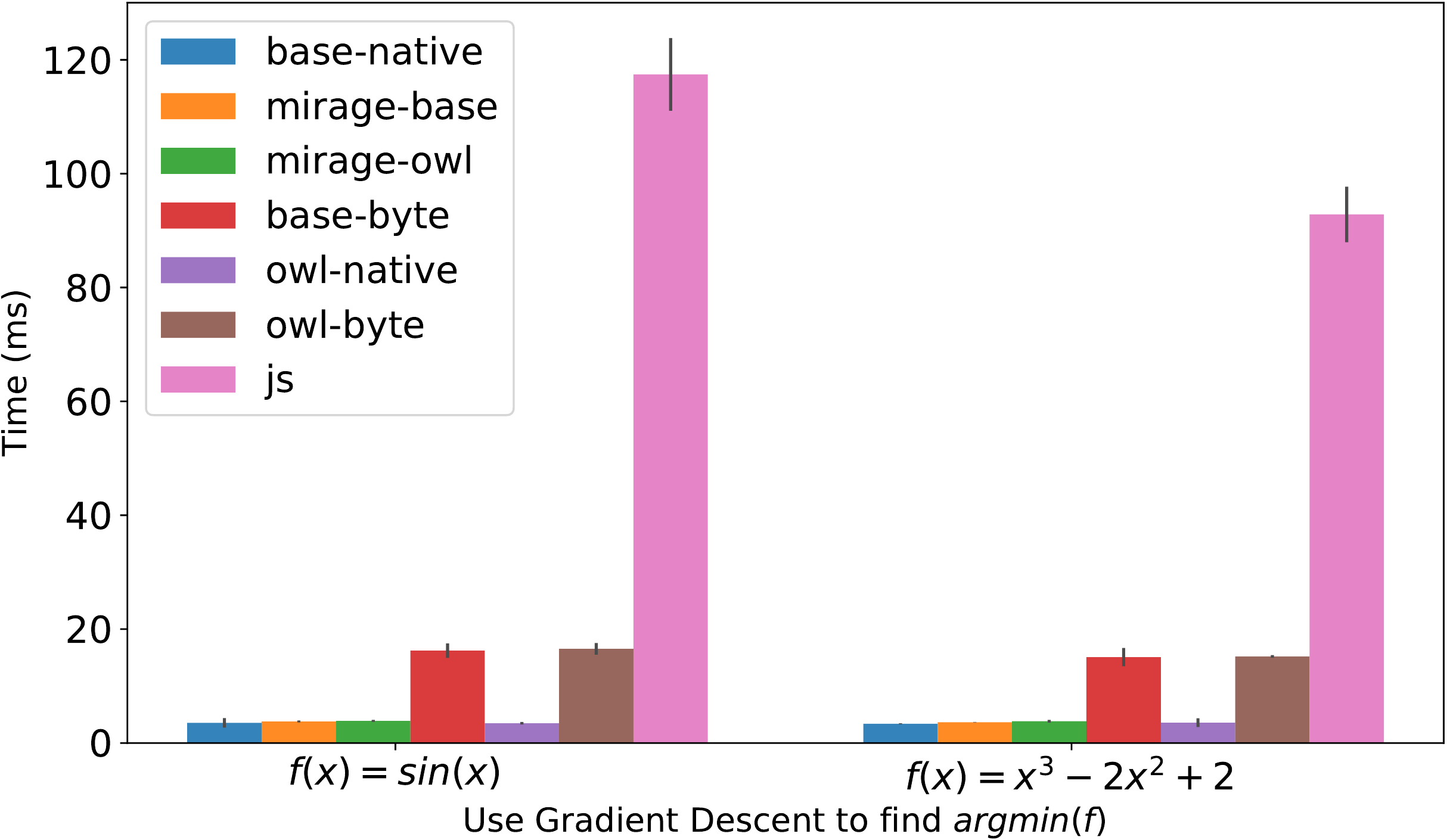}
	\caption{Performance of gradient descent on function $f$ to find $argmin(f)$ on RaspberryPi.}
	\label{fig:eval_gd_arm}
\end{figure}

\begin{table}
	\caption{Inference Speed of DNN (RaspberryPi)}
	\label{tab:eval_dnn_arm}
	\begin{tabular}{r|ll}
		\toprule
		Time (ms) &VGG&SqueezeNet\\
		\midrule
		owl-native & 160 ($\pm$ 6) & 1435($\pm$ 5) \\
		owl-byte & 162 ($\pm$ 3) & 1550($\pm$ 9) \\
		base-native & 6420.0($\pm$ 20.0) & 117250.00 ($\pm$ 330.0)\\
		base-byte & 28830.0($\pm$ 0.1) & 514420 ($\pm$ 310.0)\\
		mirage-owl & 35.6 ($\pm$ 0.1) & 359.6($\pm$ 0.1)\\
		mirage-base & 6615.9 ($\pm$ 3.0) & 118340.8 ($\pm$ 102.6)\\
		JavaScript & 31500.5($\pm$ 5.5) & 558871.0 ($\pm$ 3072.0)\\
		\bottomrule
	\end{tabular}
\end{table}

We have also conducted the same evaluation experiments on RaspberryPi 3 Model B. Fig.~\ref{fig:eval_mapfold}(c) shows the performance of fold operation on ndarray.
Besides the fact that all backends runs about one order of magnitude slower than that on the laptop, previous observations still hold.
This figure also implies that, on resource-limited devices such as RaspberryPi, the key difference is between native code and bytecode, instead of \texttt{owl-lib} and \texttt{base-lib} for this operation.
Similar also applies to the gradient descent algorithm in Fig.~\ref{fig:eval_gd_arm}, and the neural network inference in Table.~\ref{tab:eval_dnn_arm} on RaspberryPi.

\begin{table}
	\caption{Size of executables generated by backends}
	\label{tab:eval_size}
	\centering
	\begin{tabular}{r|llll}
		\toprule
		Size (KB) & native & bytecode & Mirage & JavaScript \\
		\midrule
		base & 2,437 & 4,298 & 4,602 & 739 \\
		native & 14,875 & 13,102 & 16,987 & - \\
		\bottomrule
	\end{tabular}
	\vspace{-1em}
\end{table}

Finally, we briefly compare the size of executables generated by different backends. We take the SqueezeNet for example, and the results are shown in Table~\ref{tab:eval_size}.
It can be seen that \texttt{owl-lib} executives have larger size compared to \texttt{base-lib} ones, and JavaScript code has the smallest file size.

In summary, there does not exist a dominant method of deployment for all these backends. It is thus imperative to choose suitable backend according to deployment environment.

\section{Related Work}

Moving ML analytics from cloud to edge devices faces many challenges.
One widely recognised challenge is that, compared with resource-rich computing clusters, edge and mobile devices only have quite limited computation power and working memory.
To accommodate heavy ML computation on edge devices, one solution is to train suitable small models to do inference on mobile devices~\cite{chun2009augmented}. This method leads to unsatisfactory accuracy and user experience.
Some techniques~\cite{lei2013accurate, chen2015compressing, HowardZCKWWAA17} are proposed to enhance this method.


Another challenge is to personalise analytics models.
One of our previous research work~\cite{rodriguezWZMH17} explores training personalised model on local devices from an initial shared model.
Instead of moving data from user to cloud, our method provides for model training and inference in a system where computation is moved to the data.
Specifically, we take an initial model learnt from a small set of users and retrain it locally using data from a single user.
It is proved to both be robust against adversarial attacks and can improve accuracy.


There exist several work on deployment of data analytics services.
Clipper~\cite{crankshaw2017clipper} is a general-purpose low-latency prediction serving system. It provides end users with a series of ML applications including computer vision, speech recognition, recommendation, etc. Clipper tries to maximise accuracy and throughput given certain latency budget.
However, the service or model deployment here is only limited to server-side, and the users cannot deploy their own service freely.
TensorFlow Serving~\cite{tfserving2017} tries to simplify the deployment models that are created and trained by TensorFlow.
It is similar to Zoo in its mechanism of serving a model for request from users.
However, it does not support type-safe service composing, nor does it offer flexible cross platform automatic deployment solutions using multiple backends.
Some deployment systems are limited to certain applications, such as Linear Regression model in LASER~\cite{agarwal2014laser} system, and video analytics model in NoScope~\cite{kang2017noscope}.
Serverless Architectures such as AWS Lambda~\cite{awslambda2017} allow users to deploy functions cost-efficiently. Existing serverless frameworks all bound closely with cloud computing platforms such as Amazon Web Services and Google Cloud Platform.



\section{Conclusions}

In this work we identify two challenges of conducting data analytics on edge:
service composition and deployment.
We propose the Zoo system to address these two challenges.
For the first one, it provides a simple DSL to enable easy and type-safe composition of different advanced services.
We present a use case to show the expressiveness of the code.
For the second, to accommodate the heterogeneous edge deployment environment, we utilise multiple backends, including Docker container, JavaScript, and MirageOS.
We thoroughly evaluate the performance of different backends using three representative groups of numerical operations as workload.
The results show that no single deployment backend is preferable to the others, so deploying data analytics services requires choosing suitable backend according to the deployment environment.

\section*{Acknowledgment}
	This work is funded in part by the EPSRC Databox project (EP/N028260/1), NaaS (EP/K031724/2) and Contrive (EP/N028422/1).

\bibliographystyle{IEEEtran}
\bibliography{reference}

\end{document}